# CAMERA UPDATE FOR GONG REFURBISHMENT: DEVELOPMENT AND VALIDATION

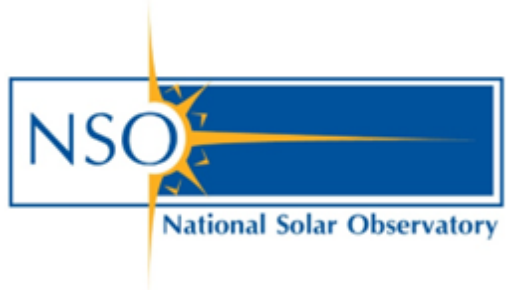

Anna L.H. Hughes, Timothy J. Purdy, Thomas M. Wentzel, Niles Oien, Luca Bertello, Sushant Tripathy, Shukur Kholikov, Kiran Jain, Gordon Petrie, Detrick D. Branston, Sanjay Gosain, Alexei Pevtsov

Institution: National Solar Observatory

January 12, 2023

Technical Report No. **NSO/NISP-2022-004**

## Abstract:

This report provides a brief summary of the properties of new cameras selected for NSF's Global Oscillations Network Group (GONG) facilities operated by the NSO Integrated Synoptic Program (NISP). These camera replacements are part of a GONG refurbishment project aimed to extend GONG operations through roughly FY 2030. Testing has confirmed the suitability of the new cameras and that current data products would be largely unchanged. GONG magnetograms show approximately one-to-one scaling with old data, and the helioseismology data (*l-nu* diagrams) are nearly identical without any identifiable artifacts. A number of tests were also conducted for GONG processing pipelines and have demonstrated that the modified NISP data center pipelines can transition smoothly to processing observations taken with the new cameras.

# 1. Outline of GONG Refurbishment

In 2016, the NSF allocated $2.5M to NSO for a multi-year refurbishment of GONG so it can continue to operate for another 10 to 15 years while its replacement (provisionally called next-generation GONG, or ngGONG) is developed. The refurbishment included upgrades to the main instrument and other non-optical but critical station components. NISP acquired a new set of cameras which will be deployed starting in FY 2023. The existing detectors are approaching their expected end-of-lifetime, making this upgrade essential in extending the network lifetime an additional five to ten years. The initial search conducted in 2017-2018 included several cameras (including one that was unexpectedly discontinued prior to purchase). Later in 2018, Emergent HB-1800-S was identified as the leading candidate, and the first test camera was acquired. Later, it was found that the test camera had manufacturing issues, and in FY 2020, a new test camera was purchased. The technical specifications, the selection criterion for the replacement camera, and detailed testing of the camera performance are described in a separate NSO technical report (Gosain et al., 2022). Following exhaustive testing of the candidate camera, the selection was finalized and an additional 11 cameras were ordered. The cameras went through acceptance tests at the NSO optical lab, which indicated the presence of strong fringes. As a potential mitigation, the entrance window in one camera was replaced by a glass window with an antireflective (AR) coating. Based on testing, which showed the AR coating on the camera's protective window reduces the amplitude of fringe patterns from 30% to about 5%, the protective windows in all cameras were replaced by windows with the AR coating. The cameras again went through acceptance testing in the optical lab (windows in two cameras failed, and later were replaced by the vendor).

Cameras to be tested were deployed to the GONG/TC engineering station in Boulder (one camera at any given time) and regular (engineering quality) data was taken, with an early focus on just one or two cameras during the long development period. Within this period there were several runs of TC observations collected for science-quality verification purposes and to test the modifications to the pipeline codes that were required for ingestion of new camera data. These data runs included a 1-month span of data in late 2021 and a more fragmented set of 64 days taken in 2022, separated in part due to trouble with the testing of camera 2, where the observations exhibited occasional abrupt changes in the Doppler velocity measurements. These changes were later associated with camera frames being dropped out. The cause of these frame dropouts was tracked to the camera (manufacturer) software. After extensive investigation, the software was updated and the issue resolved. Otherwise, no issues were identified, which confirmed that the selected cameras meet requirements.

The replacement of the cameras also requires modifications of the camera mounts, and the development of a new Data Acquisition System (DAS). The details of these two modifications will be included in a separate report.

When selecting replacement cameras, the main criteria was to minimize the impact on the main data products, which translates to a close matching of the image size (number of pixels per full disk image or pixel size in arcseconds) and a comparable readout rate, wavelength sensitivity and signal-to-noise ratio (SNR). To preserve the arc-sec-per-pixel plate scale, a small reimaging optical lens was added as part of the new camera mounts. At the time of final selection, a faster readout rate was identified as beneficial to enable "freezing out" of atmospheric seeing effects.The currently adopted camera readout rate is 180 Hz, and to achieve 60 Hz (current cameras in operation), 3 frames are added together per Doppler phase modulation state. No frame co-alignment is performed prior to averaging which is no different from the previous system.

## 1.1 Quick Comparison of Old vs. New Cameras

The optical design of GONG reimaging optics was modified such that the new camera maintains the same image characteristics as the old SMD camera. Parameters such as pixel samples across the solar full disk, i.e., plate-scale of 2.5 arc-sec per pixel (nominal) was maintained in accordance with old SMD camera images of the Sun. Further, the new EVT camera has global shutter operations and does not require smear correction.

*Table 1: Basic characteristics of SMD (old) and EVT (new) GONG cameras (for a detailed comparison, see Gosain et al, 2022).*

| Parameter | SMD Cameras | EVT Cameras |
|---|---|---|
| Camera array dimensions | 1024 x 1024 | 1600 x 1200 |
| Processed image-frame dimensions | 860 x 860 (nominal) | 860 x 860 (nominal) |
| Pixel size | 14 x 14 micron | 9 x 9 micron |
| Solar-image radius (at perihelion) | 384 pixels | 390 pixels |
| Pixel well depth | 200 Ke- | 100 Ke- |
| Frame rate | 60 fps | 180 fps (capability 240 fps) |
| Change in raw-image intensity | N/A | x 4.5 SMD |
| Change in processed continuum-image intensity | N/A | x 5 SMD |
| DAS generation | 3rd generation | 4th generation |
| Camera channels | 4 (frame transfer) | 1 (global shutter) |
| Clamp correction | yes | no |
| Smear correction | yes | no |

## 1.2 Identifying New-camera Data

The deployment of the first EVT camera to a GONG-network site (Big Bear, BB) is planned to occur in early 2023, while the deployment of EVT cameras to all six network sites will occur by the end of 2023. GONG calibration pipelines require a minimum of 30 days between camera deployments, while regular work and scheduling constraints, not to mention debugging of operational procedures following the earliest deployments, favor a schedule closer to two months between camera deployments. Therefore, we expect about a year of GONG operations during which network observations will consist of a combination of observations taken with both SMD and EVT camera systems.

We expect this transition to the new cameras to be largely transparent to our end data users. Unlike the transition to the GONG+ cameras in 2001, the new EVT cameras are meant as direct replacements to the current cameras, incurring no changes to the list or schedule of GONG output products. In particular, the near-real-time (NRT) product stream, including magnetic synoptic maps and model products should be no more affected by this transition than is usually the case when a GONG site undergoes regular, preventative maintenance (PM). (Note: for farside maps, it is expected that we will need to derive new relations between phase shift and magnetic flux that will be used in the farside pipeline. It is not clear how the probability of features (sunspots) detected on farside images will be affected during the transition/deployment period when the data being taken with mixed (old & new) cameras are merged.)

For those users who would still like to be able to track the camera source(s) of a specific data set, Table 2 provides a list of pertinent FITS header keywords that have been added to several of the GONG output products in (primarily) our synoptic-map pipelines in anticipation of the EVT camera deployment. A full list of all new header keywords associated with the DAS-4th-generation pipeline upgrade will be provided with the full report.

*Table 2: FITS header keywords added to GONG output products based on the EVT camera observations.*

| Data Type | Keyword | Definition |
|---|---|---|
| level-1 obs | EXPTIME | Camera-frame exposure time (s) |
|  | FRAMRATE | Camera rate (frames-per-second) |
| level-1 & level-2 obs | SN_CAM | Camera serial number |
|  | **CAM_TYPE** | Camera manufacturer type ('SMD' or 'EVT') |
| Synoptic maps & site-merged angle-calibrated magnetograms, remaps and velocities | SN_CAMS | List of camera serial numbers for all contributing observations |
|  | **CAM_TYPE** | Camera manufacture type for all contributing observations ('SMD', 'EVT', or 'mixed') |
|  | CAMTP_<SS><br><br>where <SS> = BB[1], CT, LE, ML, TD, UD, TE, or TC | Site-specific camera type ('SMD', 'EVT', or 'mixed') for the indicated site <SS>. (TE & TC are engineering sites.)<br><br>(These header entries are only present if **CAM_TYPE = 'mixed'** and observations from a given site <SS> are included in the synoptic map.) |

For FITS data with which these keywords do not appear, users should assume an SMD camera origin, as long as these keywords do appear in a more recent example of the specific GONG product type. These keywords have been added to the data-processing stream of all planned product types as of mid-September 2022.

---

[1] BB - Big Bear, CT - Cerro Tololo, LE - Learmonth, ML - Mauna Loa, TD - Teide, UD - Udaipur, TE and TC - engineering sites in Boulder, CO; see https://gong.nso.edu/sites/

## 1.3 Science Data Products Summary

Note that the Boulder engineering site has two telescopes. The new EVT cameras were mounted on the TC telescope, while the TE telescope continued to run with the old SMD camera.

*Table 3: Details of tests that were performed as part of the science data product verification.*

| Area | Test/Issue |
|---|---|
| Basic Camera Performance | |
| Drift-scans | Vignetting |
| | GRASP-package adjustments: camera corrections and specifications |
| | Thresholds adjustments: intensity and image quality |
| AutoVMBI | UCOR processing: camera and optical corrections |
| | Automatic image rejection (AIR) statistics & thresholds adjustments |
| Copipe | XOFFSET tests: cross-correlation between paired images for network alignment |
| Mode frequencies | with CAM_TYPE: new FITS-header keyword |
| | TC vs BB data: Pmode time series, global mode frequencies, ring-diagram data products for localized regions |
| IMMERGE | Single-site high-resolution Dopplergram (vzi) analysis including TE data |
| | Paired merged images |
| qr_3map (Air Force cubes) | TC cubes test generation |
| Farside | Low-resolution Dopplergram (fqi) comparisons |
| | Comparison of phase-shift maps |
| Zeropoint | Magnetogram inspection: spatial values and distribution comparisons |
| | Aggregate heliographic magnetograms (hamburger) inspection: magnetic zero-bias performance comparison |
| magmap(z) | Single-site synoptic maps including TC |
| Other Science Testing | Time-distance test: power spectrum and cross-correlation function |

## 1.3.1 Magnetogram Data Products

A number of tests have been performed to evaluate the quality of the new-camera 10-minute averaged magnetograms. The tests included a comparison of the weak-field distribution of the magnetic field acquired almost simultaneously at TC (new camera) and TE (current camera), and a pixel-to-pixel analysis to determine the scaling between the two sets of magnetograms. In total, 64 days of TC and TE pairs of observations taken mostly during the months of June and July 2022 have been investigated. For each full-disk magnetogram, only pixels with values less than 20 Gauss were selected, and the corresponding magnetic flux density distribution was accurately modeled using a Moffat probability distribution. Several parameter time series were then derived from these distributions, but mostly the Full-Width-Half-Maximum (FWHM) was used in this comparative study. In addition, the median absolute deviation (MAD) was also computed using only values from a 201x201 pixels square-region at the disk image center mostly dominated by quiet Sun (weak magnetic field). The FWHM and MAD time series show very similar behavior. The typical value of the FWHM is around 5 Gauss for observations taken during the hours around local noon but it is generally higher by a factor of 2, and sometimes more, at the beginning and the end of the observing day. The MAD values are slightly above 2 Gauss, and they show a similar daily trend as the FWHM data. Analysis of the FWHM and MAD time series indicates that the noise level of the magnetograms produced by the new camera is slightly lower (about 0.2 Gauss) than the noise level of the magnetograms obtained with the current camera installed at the GONG Boulder site. An overall noise level of about 2.2 Gauss for the new-camera magnetograms was also determined from the analysis of the MAD time series. This value includes contributions mostly from solar noise, atmospheric seeing, and the actual camera noise.

We have also performed a pixel-to-pixel comparison of co-temporal TE and TC magnetograms to compute the magnetic field scaling between the data acquired by the two cameras. For this purpose, we have used only pixels located in a small square region (200x200 pixels Region-Of-Interest, ROI) centered at the disk center of the solar image. The TC and TE ROIs are then properly co-aligned via a 2-D cross-correlation before generating the scattered plot. The result is a very linear relationship between the TE and TC magnetic field values, which allows one to compute the slope (scaling factor) with a high degree of confidence. We found that the derived magnetic flux density values from the new camera are about the same as those from the current camera (1:1 scaling) for most of the days examined here. The 64 days of observations used for this analysis included a period when the frame-drop issue with the camera software was being resolved. While this issue created artifacts in the Dopplergram images, our analysis shows that the artifacts in the Dopplergrams did not affect the quality of the magnetograms. Single site (TC) synoptic maps were also generated and compared with network-merged maps. A comparison of synoptic maps and Potential-field Source-surface (PFSS) models from TC and the other GONG sites for Carrington Rotation 2247 showed a good agreement.

## 1.3.2 Helioseismology Data Products

The results of tests described in this section correspond to computation of daily *l-nu* diagrams, power spectra and mode frequencies, farside and other science testing as listed in Table 3.

**Daily l-nu diagrams:** Characteristic *l-nu* (harmonic degree-frequency) diagrams were made from daily Doppler observations taken at the engineering site when the duty cycle exceeded a threshold value of 70%. These were compared with other site data, specifically with Big Bear (BB) data which is the nearest GONG network site. The comparison of power at different *l*-values illustrates excellent agreement between the old and new cameras. Since the seeing conditions are different between TC and BB sites, comparisons of *l-nu* diagrams were also made between TC and TE sites (located adjacently) and found to be consistent.

***Power Spectra and p-mode frequencies***: Doppler observations over a period of 19 days were used to compute power spectra and oscillation frequencies using the GONG pipeline. For this task, two time-series were created. The first one used data from the regular GONG sites. In the second time series, BB data was replaced by the new camera data from the TC site. This resulted in a slightly lower duty cycle (about 3%) but the fitted frequencies were found to be in agreement within the uncertainties. As the final test, the three-dimensional power spectra (two spatial and one temporal) were computed using merged data from two sites. The modes fitted using the GONG ring-diagram pipeline were found to be similar.

***Farside Mapping***: We also tested the farside pipeline using the Dopplergrams obtained from the new camera. During the process, a substantive difference in the values of the phase shifts was noticed between the new and current cameras. This resulted in the incorrect mapping of the probability values. Thus it was concluded that a new calibration for the farside pipeline will be required after the current cameras have been upgraded at all six sites. With images from 1 new camera included in the farside pipeline, we noticed changes in the phase shift values. In test cases, when we replaced BB or CT images by TC images, we considered simultaneous observations for a particular pair, e.g., BB and TC. This allowed us to compute farside maps for same duty cycle values with identical gap patterns. Since phase shifts are the basis for computing probabilities and the area, it is expected that there will be deviations from the current values when new cameras are deployed to all six sites. As a result, we need to derive new relations that will be used in the farside pipeline.

***Time-distance analysis:*** For time-distance analysis, 11 days of TC and TE parallel observations, when the duration of the observations was longer than nine hours, were selected. An average Dopplergram was created from the selected 11 days of observations and is subtracted to remove solar rotation. Daily observations were remapped into heliographic coordinates and tracked relative to the noon time. The power spectra were found to be clean enough except for a non-significant low-*l* noise.

A cross-correlation function (CCF) was computed using the same 11 selected daily 9-hour-long observations. To cover large travel distances, only 30x30 degree patches at the solar disk center were used and measurements within the patch points are averaged. CCF is found to look reasonable and clean, without any significant temporal or spatial noises. Multiple skip signals up to 7th reflections are visible.

For the alternative analysis of spatial and frequency differences, 3 consecutive days of simultaneous Dopplergrams of TC and TE were used. The results are in very good agreement, at least for such low-resolution one-day spectra. A power distribution as a function of *l* shows a non-significant difference (at most ranges in both frequency and space). Power from TC is a little bit lower at low-*l* and higher at high-*l* values.

# 1.4 New-Camera Transition Summary

A detailed report on the camera and science data-product testing will be published as an NSO Technical Report, which will also include this Executive Summary. The main highlights of this work, in terms of science, operations, and engineering, are that:

(1) The selected cameras satisfy the requirements on their sensitivity, linearity in response, and uniformity across the field of view. The science quality of all data products is largely similar to the current (SMD) cameras in operation with near one-to-one scaling for the derived magnetic field (flux density per pixel in full disk magnetograms), but a slightly higher noise level. *l-nu* diagrams do not exhibit any notable abnormalities, and limited tests of merging data to the network data product do not indicate any issues.

(2) Test data taken at our engineering site were run through the GONG data-processing pipelines from level-0 up through processing sufficient to verify that the EVT-camera changes to image intensity, image radius, and the new FITS header keywords will not impact the data processing or data output standards. Where needed, adjustments were made to the code base to handle, e.g., the new EVT-camera readout style (no clamp correction or channel boundaries), and the change in image-capture timing, though the changes in timing did present significant technical challenges.

(3) Individual EVT camera testing will continue with the remaining 11 cameras. To ensure the camera deployment to the entire network in FY 2023, the period of testing will be reduced to approximately 1-2 weeks at the GONG/TC facility. Each camera will be tested with its own camera mount and the DAS. The deployment will start after at least 2 camera-DAS sets have passed the testing. The tests will not include a complete verification of all science data products although a limited number of *l-nu* diagrams may be produced. Camera deployment will start from GONG/BB followed by GONG/Cerro Tololo (CT) or GONG/Mauna Loa (ML). These first deployments will help finalize the deployment procedures prior to traveling to the remaining international sites.